# Time-Dependent Dielectric Breakdown on Subnanometer EOT nMOS FinFETs

Pedro C. Feijoo, Thomas Kauerauf, María Toledano-Luque, Mitsuhiro Togo, Enrique San Andrés, and Guido Groeseneken,

P. C. Feijoo and M. Toledano-Luque are with the Departamento de Física Aplicada III, Universidad Complutense de Madrid, 28040 Madrid, Spain, and also with IMEC, 3001 Leuven, Belgium (e-mail: pedronska@fis.ucm.es).

T. Kauerauf, M. Togo, and G. Groeseneken are with IMEC, 3001 Leuven, Belgium.

E. San Andrés is with the Departamento de Física Aplicada III, Universidad Complutense de Madrid, 28040 Madrid, Spain.

**Abstract**

In this paper, the time-dependent dielectric breakdown (TDDB) in sub-1-nm equivalent oxide thickness (EOT) n-type bulk FinFETs is studied. The gate stacks consist of an IMEC clean interfacial layer, atomic layer deposition $HfO_2$ high-κ and TiN metal electrode. For the 0.8-nm EOT FinFETs, it is found that TDDB lifetime is consistent with results of planar devices for areas around $10^{-8}$ $cm^2$, implying that the FinFET architecture does not seem to introduce new failure mechanisms. However, for devices with smaller area, the extrapolated voltage at a ten-year lifetime for soft breakdown (SBD) does not meet the specifications, and as a consequence, the SBD path wear-out will have to be included in the final extrapolation. Furthermore, it is shown that for EOTs smaller than 0.8 nm, the TDDB reliability on n-type FinFETs is challenged by the high leakage currents.

I. **INTRODUCTION**

Multigate field-effect transistors (MuGFETs or FinFETs) are one of the strongest contenders to replace planar CMOS in the near future. In fact, Intel has announced that its 22-nm process will be a triple-gate transistor. They present better short-channel behavior and potential area benefits as compared with their bulk planar counterparts [1]–[3]. In combination with the small dimensions of the devices, higher drive current and lower off current are obtained. The outcome of these benefits is better device performance and less power consumption.

FinFETs consist of silicon fins that connect the source and the drain, partially surrounded by the gate stack. Depending on the number of sides of the silicon fin that are covered by the gate stack, FinFETs are classified into double-, triple-, or quadruplegate FETs [4].

Additionally, to meet the sub-32-nm roadmap requirements, the equivalent oxide thickness (EOT) has to be reduced below 1 nm [5]. Gate stacks based on high-$\kappa$ hafnium dioxide ($HfO_2$) and TiN as metal electrode are used in order to get these low EOTs, since the TiN gate is known to scavenge the interfacial silicon oxide that unavoidably grows between the silicon and the high-$\kappa$ [6]. The EOT is controlled by the metal gate thickness, which modulates the scavenging effect.

Therefore, a reliability study on FinFET devices with EOTs below 1 nm is essential. In this paper, time-dependent dielectric breakdown (TDDB) is assessed on triple-gate bulk FinFETs with a high-$\kappa$/TiN gate stack. EOTs range from 0.8 to 0.7 nm. Bulk FinFET fabrication flow is described in [7].

II. **EXPERIMENTAL**

In Fig. 1, the FinFET device structure used in this work is described. The silicon fin is directly connected to the bulk, and the top and lateral surfaces of the fin body are wrapped by the gate dielectric and metal electrode. This triple-gated structure presents a good compromise between performance and process integration complexity [1], [8], as compared with other FinFET architectures. From its 3-D structure, the width ($W_{fin}$) and the height ($H_{fin}$) of the fin are defined as shown in Fig. 1(a). The distance from the gate edge to the source and the drain is called fin extension ($L_{ext}$), as in Fig. 1(b). A device can contain one or multiple fins, hence the number of fins ($N_{fin}$) and the distance between two adjacent fins or pitch ($S$) are parameters that have to be taken into account. Thus, the total device area $A$ of the channel follows this expression:

$$A = (2H_{fin} + W_{fin})L_G N_{fin}. \qquad (1)$$

The n-type FinFETs were fabricated on 300-mm (100) Si wafers. HfO$_2$ was deposited by atomic layer deposition. To minimize EOT, a thin TiN/Si-cap gate structure was used to scavenge the interfacial layer. TiN was deposited by physical vapor deposition. In order to achieve gate stacks with different EOTs, metal gate thicknesses of 5 and 3 nm were used. Fig. 2 shows a transmission electron microscopy (TEM) image of the structure with 5-nm-thick TiN. The reduction in the interfacial layer is more effective for the thinner TiN layers and this resulted in EOTs of 0.7 nm for the 3-nm TiN and 0.8 nm for the 5-nm TiN.

The number of fins ($N_{fin}$) is 5, and devices with a gate length $L_G$ of 35 nm and 1 μm are compared. Fin width ($W_{fin}$), fin height ($H_{fin}$), and distance between fins ($S$) are 20, 27, and 300 nm, respectively, for the evaluated devices.

The TDDB test is performed at 125 °C using a constant voltage stress (CVS), where a gate voltage ($V_G$) is applied with the source, drain, and bulk grounded. The gate leakage current was continuously measured during stress. In order to extract the time-to-breakdown ($t_{BD}$) for each device, it is necessary to define a failure criterion, which should distinguish between soft breakdown (SBD) and hard breakdown (HBD) [9]. Assuming that $t_{BD}$ follows a Weibull distribution, from the TDDB data, we can fit the Weibull slope $β$ using the maximumlikelihood estimation. Considering SBD, for an intrinsic failure distribution, $β$ is related to the number of traps that form a percolation path through the dielectric [10]. The time to failure of 63% of devices ($η$) as a function of stress voltage can be also calculated. The maximum voltage for a ten-year lifetime can be extrapolated using a power law model [11]. Thus

$$t_{BD} = B \cdot V_G^{-n} \qquad (2)$$

where $B$ is a constant, and $n$ is the acceleration factor. To obtain the lifetime according to the TDDB reliability specification, this extrapolation must be scaled to a gate dielectric area of 0.1 cm² and 0.01% failures. For the next generation of electronic devices, the International Technology Roadmap for Semiconductor projects a supply voltage of 0.9 V [5]. Then, it should be assured that transistors correctly operate for ten years withstanding this voltage.

### III. RESULTS AND DISCUSSION

In Fig. 3, TDDB results for the 0.8-nm EOT FinFET (5-nm TiN) with $L_G$ = 1 μm and $W_{fin}$ = 20 nm are shown. The area of these devices is $3.70 \times 10^{-9}$ cm², which is too large to clearly observe individual SBD paths. The leakage current of around 10 μA is significantly higher compared with the current through single paths created by stress (~10– 100 nA) [12], whereas the creation of multiple parallel paths leads to a progressive increase in the total current or stressinduced leakage current (SILC) instead of showing an abrupt increase. Only when one of these SBD paths suffers from wearout, HBD occurs [10] as shown in Fig. 3(a). The time-to-hard breakdown ($t_{HBD}$) is obtained using a current step trigger of 10 μA, and the Weibull distributions for different gate voltages are represented in Fig. 3(b). The Weibull slope $\beta_{HBD}$ is 1.53, and in Fig. 3(c), the lifetime is extrapolated, obtaining a maximum allowable HBD voltage of 1.3 V for a ten-year lifetime and an acceleration factor $n$ of 56.

If these obtained parameters are compared with the results for planar nMOS devices with similar area and EOT [13], it is found that the values are the same within error (for planar devices, $\beta_{HBD}$ is 1.54, the acceleration factor is 47, and the projected allowable voltage is over 1.2 V). This means that the FinFET structure does not introduce new mechanisms of CVS degradation in nMOS. Moreover, leakage current densities are very similar for the same stress voltages ($10^3$ to $5 \times 10^3$ A · cm⁻² at $V_G$ = 2.2−2.5 V). One of the major concerns in FinFETs is the concentration of electric field around the fincorners, which may induce preferential dielectric breakdown in these regions. Then, the result that nMOS TDDB measurements do not change for FinFET architecture implies that corners do not affect breakdown in this case.

Current–time traces for devices with $L_G$ = 35 nm and $W_{fin}$ = 20 nm ($A$ = 1.30 × 10$^{-10}$ cm$^2$) are depicted in Fig. 4. For this smaller area, the gate current and the probability to create a leakage path is lower than in the larger area case. As a consequence, after the beginning of the stress, the gate current is constant for a longer time, and the creation of a single percolation path is easily recognizable. Thus, SBD can be investigated. Using a current step breakdown trigger of 100 nA, the time-to-soft breakdown ($t_{SBD}$) is extracted for each device, and the statistics are obtained for the different stress voltages. A Weibull slope $\beta_{SBD}$ of 1.08, indicating that a path of three traps is necessary to reach SBD [11], and an acceleration factor $n$ of 30 are obtained. Since the extrapolated gate voltage at a ten-year lifetime is only 0.7 V (see Fig. 5), the target voltage is not met when considering SBD only and, as with low EOT planar devices, the wear-out phase before HBD must be included in the TDDB lifetime extrapolation [9], [12].

Fig. 6 represents current traces for devices with $A$ = 3.70 × 10$^{-9}$ cm$^2$ ($L_G$ = 1 μm, $W_{fin}$ = 20 nm) and EOT of 0.7 nm (TiN gate thickness of 3 nm). It can be observed that, due to the relatively low EOT and the large channel length, the initial leakage current is already high, and the immediate creation of multiple paths during stress prompts a high SILC. However, before HBD, the leakage current reaches a level where series resistance begins to affect the measurement (typically around 200 μA) [12]. Therefore, not even HBD can be measured in such devices, and for an nMOS TDDB evaluation, only smaller areas can be used.

In Fig. 7, the current–time traces of the smaller area devices ($L_G$ = 35 nm, $A$ = 1.30 × 10$^{-10}$ cm$^2$) with an EOT of 0.7 nm are represented. The gate current is lower, and a current step trigger of 200 nA was used to obtain the time-to-soft breakdown ($t_{SBD}$). In this case, the Weibull slope is 1.15, indicating again that SBD is formed by a three-trap path through the dielectric. The power law acceleration factor $n$ is 32, and the extrapolated voltage at a ten-year lifetime is 0.7 V (see Fig. 5). These results are similar to those of the devices with the same area and EOT of 0.8 nm. As observed for devices with 0.8-nm EOT, this extrapolated voltage does not reach the expected value for the next generation of transistors, and in future work, the wear-out has to be included in the TDDB lifetime extrapolation.

## IV. CONCLUSION

CVS was applied to n-type bulk FinFETs with sub-1-nm EOT in order to assess the TDDB reliability. It was found that the HBD results for FinFETs with similar area (∼10$^{-8}$ cm$^2$) and EOT (0.8 nm) are analogous to those of the planar devices. This indicates that, for this multigate architecture, the breakdown behavior is not affected, and no extra degradation mechanism appears in this case. However, for smaller EOT FinFETs, it is, similar to planar devices, very challenging to evaluate TDDB. This happens because the leakage currents are too high to observe SBD, and the series resistance hampers HBD measurements. For smaller channel length (and smaller area) FinFETs, the SBD TDDB reliability can be evaluated. The extrapolated voltages at a ten-year lifetime are 0.7 V, which is below the specifications. To prove sufficient TDDB reliability in future work, the wear-out before HBD must be included in the TDDB lifetime extrapolation.

## ACKNOWLEDGMENTS

The authors would like to thank IMEC's p-line for the processing and AMSIMEC for the electrical characterization. This work is part of the Interuniversity Microelectronics Center's (IMEC)

Industrial Affiliation Program, funded by IMEC's core partners, and was supported by the Spanish Ministry of Education and Ministerio de Ciencia e Innovación (MICINN) under FPU Grant AP2007-01157 and Project TEC2010-18051.

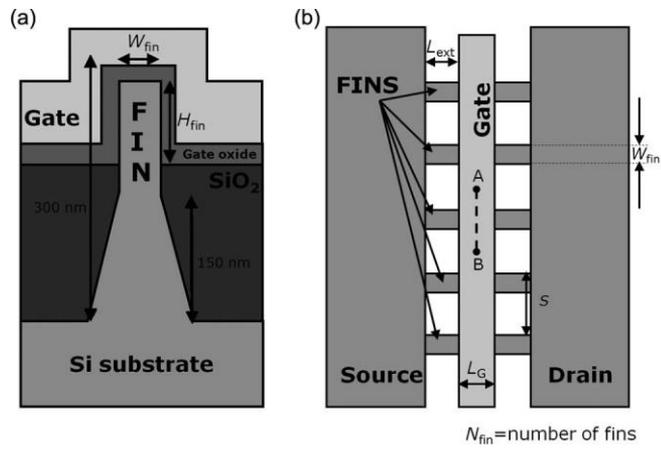

Fig. 1. (a) Schematic cross-sectional view of a fin. (b) Schematic top view of a FinFET device consisting of several fins. (Cutline along A–B for fin crosssectional view.)

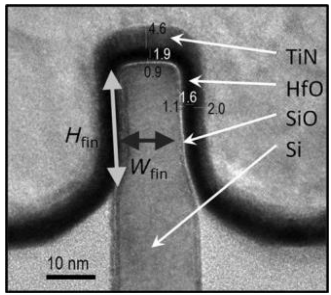

Fig. 2. TEM image of a FinFET with a TiN/HfO$_2$ stack with a TiN layer with 5 nm of target thickness.

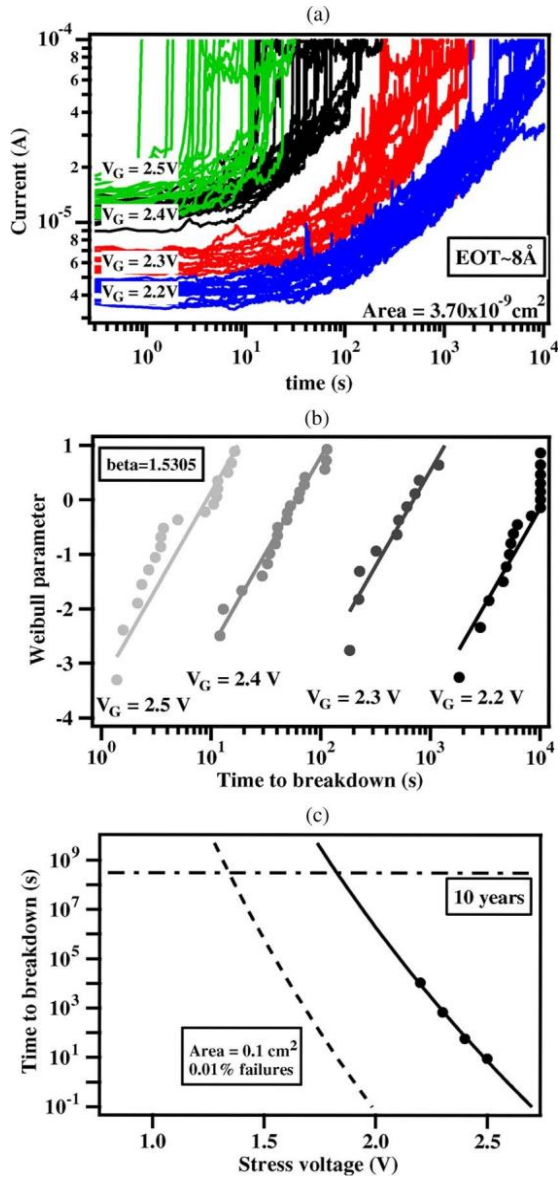

Fig. 3. TDDB results for n-type FinFET devices with EOT ~8 Å and A = 3.7× 10⁻⁹ cm² stressed at 125 C. (a) Current traces. (b) Weibull plot. (c) Projected lifetime. Current step for breakdown is 10 $\mu$A. Extrapolated operating voltage at a ten-year lifetime is around 1.3 V.

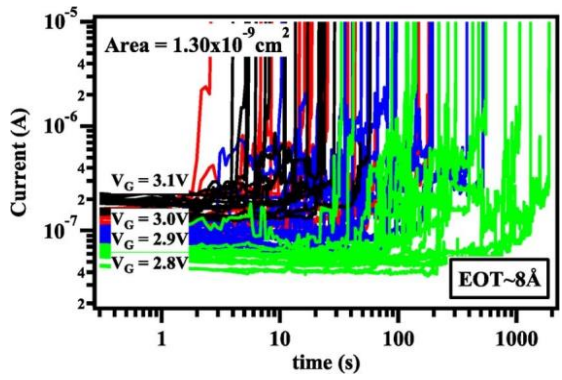

Fig. 4. Current traces for n-type FinFET devices with 8-Å EOT and $A = 1.30 \times 10^{-10}$ cm². SBD is evaluated with a current step of 100 nA.

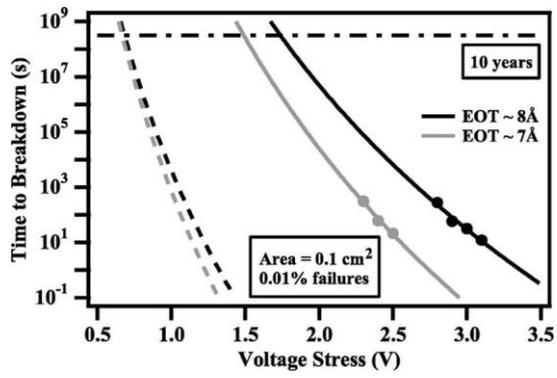

Fig. 5. Ten-year lifetime extrapolation for SBD of FinFETs with EOTs of 7 and 8 Å and $A = 1.30 \times 10^{-10}$ cm$^2$.

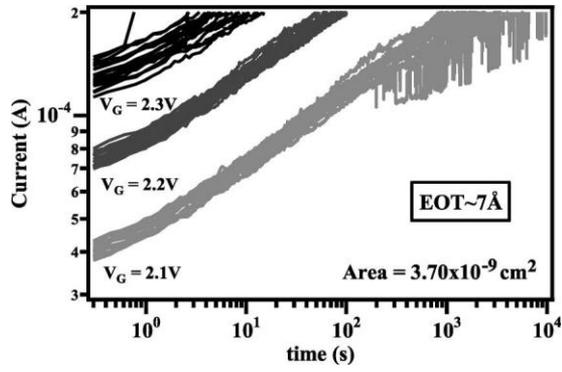

Fig. 6. Current–time curves for n-type FinFET devices with 7-Å EOT and $A = 3.70 \times 10^{-9}$ cm$^2$.

Currents reach series resistance limit before breakdown

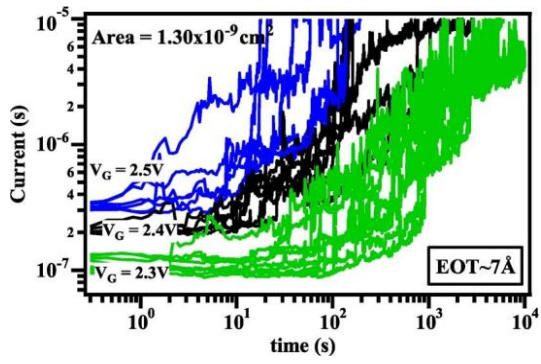

Fig. 7. Current–time curves for n-type FinFETs with 7-Å EOT and $A = 1.30 \times 10^{-10}$ cm$^2$.